\documentclass[aps,prb,twocolumn,superscriptaddress,groupedaddress]{revtex4-2}

\usepackage{amsmath,amssymb,amsfonts}
\usepackage{bm, latexsym}
\usepackage{graphicx}
\usepackage{bbm,times}
\usepackage[normalem]{ulem}
\usepackage[pdftex]{xcolor}
\usepackage[pdftex]{hyperref}
\usepackage{braket}

\usepackage{comment}

\newcommand{\cred}[1]{\textcolor{black}{#1}}

\graphicspath{
{./image/}
}

\begin{document}

\title{Phase diagram of non-Hermitian BCS superfluids in a dissipative asymmetric Hubbard model}
\author{Soma Takemori}
\email{s-takemori@stat.phys.titech.ac.jp}
\author{Kazuki Yamamoto}
\author{Akihisa Koga}
\affiliation{Department of Physics, Tokyo Institute of Technology,
  Meguro, Tokyo 152-8551, Japan}

\date{\today}

\begin{abstract}
  We investigate the non-Hermitian (NH) attractive Fermi-Hubbard model
  with asymmetric hopping and complex-valued interactions,
  which can be realized by collective one-body loss and two-body loss.
  By means of the NH BCS theory, 
  we find that the weak asymmetry of the hopping does not affect the BCS superfluidity
  since it only affects the imaginary part of the eigenvalues of the BdG Hamiltonian.
  Systematic analysis in the $d$-dimensional hypercubic lattices
  clarifies that the singularity in the density of states
  affects the phase boundary between the normal
  and dissipation-induced superfluid states.
  Our results can be tested in ultracold atoms by using the photoassociation techniques and a nonlocal Rabi coupling with local losses and postselecting null measurement outcomes with the use of the quantum-gas microscope.
\end{abstract}

\maketitle

\section{Introduction}\label{sec_intro}
Strongly correlated electron systems have attracted broad interest due to recent development of experimental techniques
for ultracold atomic systems~\cite{schafer20}.
Due to the high controllability~\cite{inouye98,chin10,zhang15,hofer15,pagano15,cappellini19}, the ultracold atomic systems can be regarded as the platform of quantum simulations~\cite{zwierlein04,regal04,schunck08,behrle18,biss22,sobirey22,holten22}.
In fact, many-body phenomena in the optical lattice have been realized such as in Mott insulating, antiferromagnetic,
and superconducting states~\cite{chin06,jordens08,cheuk16mott,scott19,muniz20}.
Recently, open systems, where the dissipation is inevitable due to the coupling to the environment,
have attracted much attention~\cite{daley14,sieberer16,ashida20,meden23}.
This makes many-body physics more fruitful,
and interesting phenomena have been observed
such as dynamical sign reversal of the spin correlation~\cite{honda23} and the characteristic superfluid transport~\cite{huang23}.
In addition, many experiments with dissipative process due to the inelastic collision of atoms have been conducted for, e.g., the single-body loss~\cite{barontini13,patil15,labouvie16,luschen17,corman19,takasu20,bouganne20,benary22,ren22,huang23}, the two-body loss~\cite{syassen08,yan13,zhu14,tomita17,sponselee18,tomita19,honda23}, and the three-body loss~\cite{mark12,mark20}.
These stimulate further theoretical investigations of the dissipation effect on the many-body physics~\cite{ashida16,ashida17,lourencco18,nakagawa18,hanai19,hamazaki19,nakagawa20,yamamoto20,liu20,xu20,mu20,zhang20,hanai20,matsumoto20,yang21,zhang21eta,nakagawa21,tajima21,yamamoto22,zhang22,dora22,orito23,yamamoto23sun,han23,wang23,rosso23,sarkar23,yu24,yang24,yamamoto24}.

The superfluid state in open systems has been widely investigated~\cite{han09,diehl10dwave,gahtak18,yamamoto19,Damanet19,yamamoto21,he21,kanazawa21,iskin21,li23,tajima23topo,mazza23,takemori24,tajima24,shi24}.
It has been clarified that, in addition to the conventional normal and superfluid states,
the dissipation-induced (DS) superfluid state appears in the NH system,
where the continuous quantum Zeno effect (QZE) induced by strong two-body losses plays an important role~\cite{yamamoto19}. 
The effect of the asymmetric hopping, which can be realized by the nonlocal one-body loss,
has been discussed so far.
It has been clarified that fruitful phenomena are induced
by nonreciprocal effects~\cite{fukui98,hatano96,hatano97,hatano98,gong18}.
Then, a question arises; how stable is the superfluid state
against both the nonlocal one-body loss and two-body loss in the ultracold fermionic systems?

To answer this question, we consider the NH attractive Hubbard model
with complex-valued interactions and asymmetric hopping.
To discuss how stable the superfluid state is against these dissipations,
we employ the NH BCS theory and obtain the NH gap equation.
Then, we clarify that the asymmetry of the hopping has no effects on the gap equation,
and only contributes to the imaginary part of the dispersion relations.
By performing the self-consistent calculations,
we obtain the phase diagrams for the hypercubic lattice in arbitrary dimensions.
In experiments, our model can be tested in ultracold atoms by postselecting the null measurement outcomes with the use of the quantum-gas microscope.

The rest of this paper is organized as follows.
In Sec.~\ref{sec_model}, we study the NH asymmetric Hubbard Hamiltonian with a complex-valued interaction on a cubic lattice.
We clarify the effect of the asymmetric hopping on the BCS superfluidity in Sec.~\ref{sec_NHBCS_asymhop}.
Section~\ref{sec_generalization_for_d_dim} is devoted to the generalization of the results to arbitrary dimensions.
Finally, a conclusion and discussion are given in Sec.~\ref{sec_discussion}.

\section{Model}
\label{sec_model}

We first consider the dissipative dynamics of ultracold fermionic atoms,
which should be described by the following Markovian Lindblad equation~\cite{daley14}
\begin{figure}[b]
  \centering
  \includegraphics[width=7.5cm]{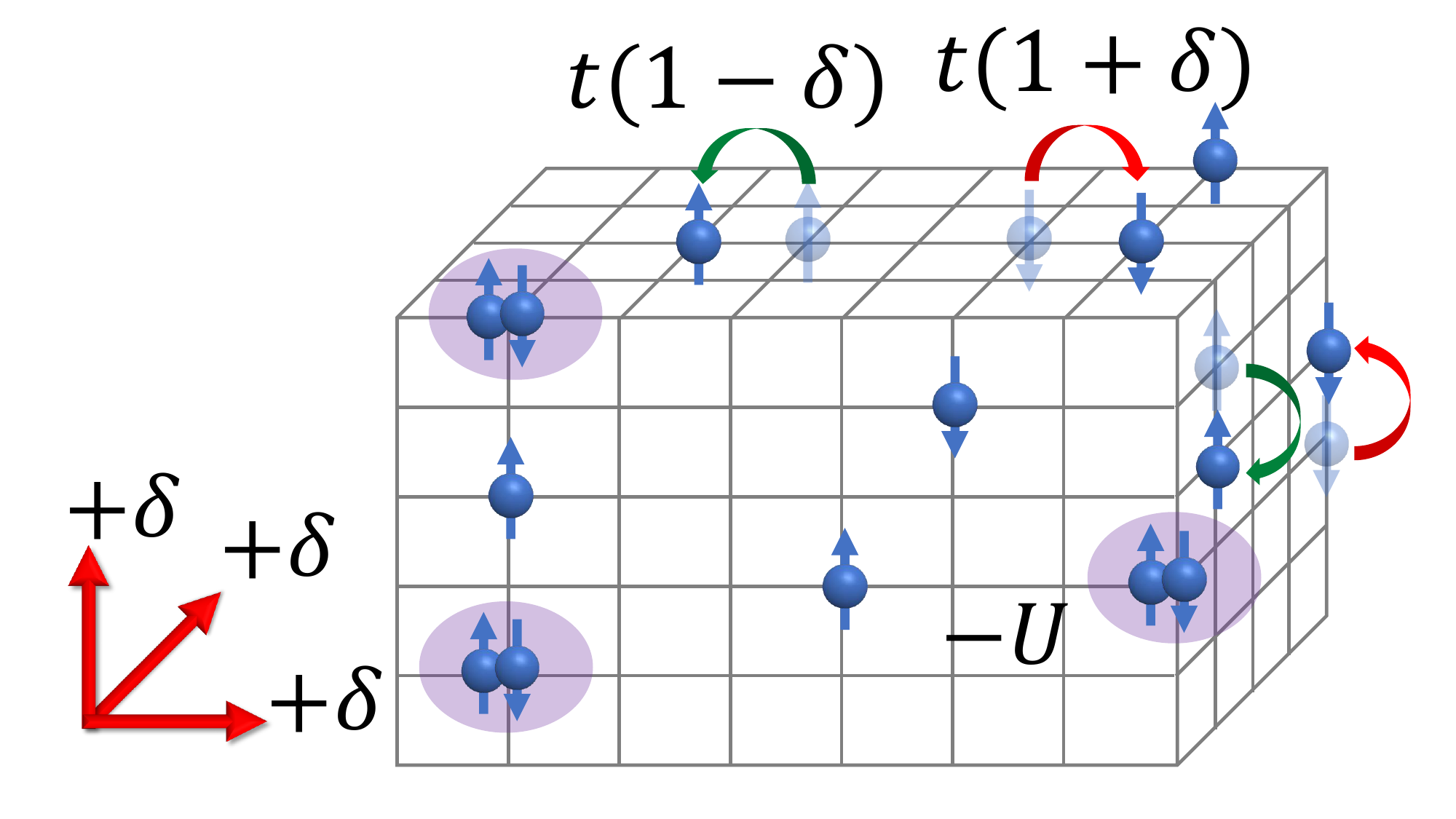}
  \caption{Schematic image of the effective NH dynamics with asymmetric hopping and the complex-valued interactions due to the two-body losses on a three-dimensional optical lattice. The hopping amplitude $t$ is modified with the asymmetry $\delta$. Here, $t(1\pm \delta)$ denotes the asymmetric hopping along the positive (negative) direction for $x,y,$ and  $z$ axes (red and green arrows). The onsite interaction $U=U_{1}+i\gamma/2$ has a complex value, where $U_{1}$ is the attractive interaction and $\gamma$ is the rate of the two-body loss.}
  \label{asymhopNHBCS_3doptlat_image}
\end{figure}
\begin{align}
   & \frac{\partial \rho}{\partial t} = -i[H_{0},\rho] + \mathcal{L}\rho, \\
   & \mathcal{L}\rho = \sum_{i}\mathcal{D}^{(i)}(\rho),
\end{align}
where $\rho$ is the density matrix, $H_0$ is the Hamiltonian of the system, and 
\begin{equation}
  \mathcal{D}^{(i)}(\rho) =\sum_{m} L_{m}^{(i)}\rho {L_{m}^{(i)}}^{\dagger} -(1/2)\{{L_{m}^{(i)}}^{\dagger}L_{m}^{(i)},\rho\},
\end{equation}
is the dissipator for the Lindblad operator $L_{m}^{(i)}$.
When two-component fermionic atoms in a three-dimensional optical lattice are considered,
the Hamiltonian $H_{0}$ is given as
\begin{align}
  H_{0} & = -t\sum_{\langle i,j\rangle,\sigma} (c_{i,\sigma}^{\dagger}c_{j,\sigma} + \text{H.c.}) - \mu\sum_{i,\sigma}n_{i,\sigma}\notag \\
  &-U\sum_{i}c_{i,\uparrow}^{\dagger}c_{i,\downarrow}^{\dagger}c_{i,\downarrow}c_{i,\uparrow}, \label{asymhopNHBCS_Hermitian_Hamiltonian_3d_Hubbard_model_eq}
\end{align}
where $c_{i,\sigma}(c_{i,\sigma}^{\dagger})$ is the annihilation (creation) operator of a fermion with spin $\sigma(=\uparrow,\downarrow)$ at site $i$, \cred{$\langle i,j\rangle$ means the summation over nearest neighbor pairs that satisfies $i<j$}, and $n_{i,\sigma}(=c_{i,\sigma}^{\dagger}c_{i,\sigma})$ is the particle number operator,
$t$ is the hopping amplitude, $\mu$ is the chemical potential, and $U_1$ is the attractive interaction.

In this study, we consider the collective one-body loss and two-body loss as dissipation in the ultracold atomic systems.
\cred{The Lindblad operator for the collective one-body loss is given by
$L^{(1)}_{i,j,\sigma}=\sqrt{\gamma_1} (c_{i,\sigma} +ic_{j,\sigma})$
with the rate $\gamma_1$.
$L^{(2)}_i=\sqrt{\gamma_2}c_{i,\downarrow}c_{i,\uparrow}$ describes the two-body loss at site $i$ with rate $\gamma_2$. The collective one-body loss is realized through a nonlocal Rabi coupling between the lattice and an additional auxiliary lattice~\cite{reiter12,gong18,liu19,he21}. The details are provided explicitly in Appendix~\ref{app_implementation_asymhop}.}
The dissipative dynamics is then described as
\begin{align}
   & \frac{\partial \rho}{\partial t} = -i\big(H_{\text{eff}}\rho - \rho H_{\text{eff}}^{\dagger} \big)  \notag                                                          \\
   & +\sum_{\langle i,j\rangle,\sigma} L^{(1)}_{i,j,\sigma}\rho {L^{(1)}_{i,j,\sigma}}^\dagger + \sum_{i} L^{(2)}_i\rho {L^{(2)}_i}^\dagger, \label{asymhopNHBCS_effective_lindblad_eq}
\end{align}
where the effective Hamiltonian is given as,
\begin{align}
  H_{\text{eff}}  &= -t\sum_{\langle i,j \rangle,\sigma} \large[(1+\delta) c_{i,\sigma}^{\dagger}c_{j,\sigma} + (1-\delta)c_{j,\sigma}^{\dagger}c_{i,\sigma}\large] \notag \\
                    & - \mu\sum_{i,\sigma}n_{i,\sigma} -U\sum_{i}c_{i,\uparrow}^{\dagger}c_{i,\downarrow}^{\dagger}c_{i,\downarrow}c_{i,\uparrow}, \label{asymhopNHBCS_effHamltonian_real_space_rep_eq}
\end{align}
where $\delta=\gamma_{1}/(2t),\gamma = \gamma_{2}$, and $U(=U_{1}+i\gamma/2)$ is the complex-valued interaction.
We have ignored the term $-i\delta \sum_{i,\sigma}n_{i,\sigma}$
since it does not affect the results qualitatively.
The nonlocal one-body loss and two-body loss lead to the direction-dependent hopping and complex-valued interactions
in the effective NH Hamiltonian $H_{\rm eff}$, respectively.
The effective NH Hamiltonian is schematically shown in Fig.~\ref{asymhopNHBCS_3doptlat_image}.

\cred{
To extract the effective dynamics of the NH Hamiltonian~\eqref{asymhopNHBCS_effHamltonian_real_space_rep_eq}, we employ the quantum trajectory method~\cite{daley14}, which is obtained by stochastically unraveling the Lindblad equation~\eqref{asymhopNHBCS_effective_lindblad_eq}. The quantum trajectory is characterized by two parts: the nonunitary Schr\"{o}dinger evolution process under the NH Hamiltonian~\eqref{asymhopNHBCS_effHamltonian_real_space_rep_eq}, and the quantum-jump process described by the Lindblad operator in the last two terms of Eq.~\eqref{asymhopNHBCS_effective_lindblad_eq}. By postselecting the special measurement outcomes where the quantum jump does not occur, the effective dynamics of the system is described by the NH Hamiltonian~\eqref{asymhopNHBCS_effHamltonian_real_space_rep_eq}.
}

By using the Fourier transformation
\begin{align}
  c_{i,\sigma} = \sqrt{\frac{1}{N}}\sum_{\bm{k}}c_{\bm{k},\sigma}e^{i\bm{k}\cdot\bm{r}_{i}},
\end{align}
the effective Hamiltonian reads
\begin{align}
  H_{\text{eff}} & = \sum_{\bm{k},\sigma} \xi_{\bm{k}}c_{\bm{k},\sigma}^{\dagger}c_{\bm{k},\sigma} - \frac{U}{N}\sum_{\bm{k},\bm{k}'}c_{\bm{k},\uparrow}^{\dagger}c_{-\bm{k},\downarrow}^{\dagger}c_{-\bm{k}',\downarrow}c_{\bm{k}',\uparrow}, \label{asymhopNHBCS_effHamltonian_momentum_space_rep_eq} \\
  \xi_{\bm{k}}   & = -2t\sum_{j=x,y,z}\cos k_{j} -2it\delta \sum_{j=x,y,z}\sin k_{j}-\mu, \label{asymhopNHBCS_energy_dispersion_3d_eq}
\end{align}
where $N$ is the number of the sites.
We find that
two kinds of dissipation make the Hamiltonian non-Hermitian, that is,
the collective one-body loss yields the imaginary part of the dispersion relations
and the two-body loss yields the imaginary part of the interactions.

\section{NH BCS theory with asymmetric hopping and a complex-valued interaction and the effect of the asymmetry of the hopping} \label{sec_NHBCS_asymhop}
Here, we deal with the NH Hamiltonian~\eqref{asymhopNHBCS_effHamltonian_momentum_space_rep_eq}
in the framework of the NH BCS theory~\cite{yamamoto19}.
The NH BCS mean-field Hamiltonian is given by
\begin{align}
  H_{\text{eff}}^{\text{BCS}} & = \sum_{\bm{k}} \bm{v}_{\bm{k}}^\dag \hat{M}_{\bm{k}} \bm{v}_{\bm{k}}+E_0, \label{asymhopNHBCS_NHBCS_effective_Hamiltonian_rep_eq} \\
  \hat{M}_{\bm{k}}            & =
  \begin{pmatrix}
    \xi_{\bm{k}} & \Delta \\ \bar{\Delta} & -\xi_{\bm{k}}^{\ast}
  \end{pmatrix},
\end{align}
where $\bm{v}_{\bm{k}}^\dag=(c_{\bm{k},\uparrow}^{\dagger} \; c_{-\bm{k},\downarrow})$,
$\Delta$ and $\bar{\Delta}$ are the superfluid order parameters, and
$E_{0}=N\Delta\bar{\Delta}/U+\sum_{\bm{k}}\xi_{\bm{k}}^{\ast}$.
To evaluate the superfluid order parameter,
we define the right (left) BCS state $|\text{BCS}\rangle_{R(L)}$ as a vacuum state of Bogoliubov quasiparticle operators as follows:
\begin{align}
   & |\text{BCS}\rangle_{R} = \prod_{\bm{k}}(u_{\bm{k}}+v_{\bm{k}}c_{\bm{k},\uparrow}^{\dagger}c_{-\bm{k},\downarrow}^{\dagger})|0\rangle, \label{asymhopNHBCS_BCS_state_Right_rep_eq}                    \\
   & |\text{BCS}\rangle_{L} = \prod_{\bm{k}}(u_{\bm{k}}^{\ast}+\bar{v}_{\bm{k}}^{\ast}c_{\bm{k},\uparrow}^{\dagger}c_{-\bm{k},\downarrow}^{\dagger})|0\rangle, \label{asymhopNHBCS_BCS_state_Left_rep_eq}
\end{align}
where $u_{\bm{k}},v_{\bm{k}},\bar{v}_{\bm{k}}$ are the coefficients and $|0\rangle$ is the vacuum for the fermion.
Then, the superfluid order parameters are given by
\begin{align}
  \Delta       & = -\frac{U}{N}\sum_{\bm{k}}{}_{L}\langle c_{-\bm{k},\downarrow}c_{\bm{k},\uparrow}\rangle_{R}, \label{asymhopNHBCS_Superfluid_orderparameter_Delta_rep_eq}                        \\
  \bar{\Delta} & = -\frac{U}{N}\sum_{\bm{k}}{}_{L}\langle c_{\bm{k},\uparrow}^{\dagger}c_{-\bm{k},\downarrow}^{\dagger}\rangle_{R}, \label{asymhopNHBCS_Superfluid_orderparameter_barDelta_rep_eq}
\end{align}
where we have introduced the NH expectation value as ${}_{L}\langle \cdot\rangle_{R}\equiv {}_{L}\langle\text{BCS}|\cdot|\text{BCS}\rangle_{R}$.
We note that these order parameters~\eqref{asymhopNHBCS_Superfluid_orderparameter_Delta_rep_eq} and \eqref{asymhopNHBCS_Superfluid_orderparameter_barDelta_rep_eq} are complex in general.
Since the energy dispersion~\eqref{asymhopNHBCS_energy_dispersion_3d_eq} satisfies
${\rm Re}\xi_{\bm{k}}={\rm Re}\xi_{-\bm{k}}$ and ${\rm Im} \xi_{\bm{k}}=-{\rm Im} \xi_{-\bm{k}}$,
the matrix $\hat{M}_{\bm{k}}$ can be divided into two terms as
\begin{align}
  \hat{M}_{\bm{k}}=
  \begin{pmatrix}
    {\rm Re} \xi_{\bm{k}} & \Delta \\ \bar{\Delta} & -{\rm Re}\xi_{\bm{k}}
  \end{pmatrix}
  +i\text{Im}\xi_{\bm{k}} \hat{I},
\end{align}
where $\hat{I}$ is the identity matrix.
We note that the imaginary part of the dispersion relation has no effects on the Bogoliubov transformation
since the latter term is represented by the identity matrix.
Then, this simply affects the imaginary part of the dispersion for the quasiparticles.
This is in contrast to the Zeeman effects in the BCS theory, the transformation of
which is represented by the identity matrix with a real coefficient.
It is known that
the magnetic field affects the dispersion relation for the quasiparticles and
the superconducting state becomes unstable~\cite{clogston62,chandrasekhar62,sheehy07}.

One can diagonalize the Hamiltonian~\eqref{asymhopNHBCS_NHBCS_effective_Hamiltonian_rep_eq} as
\begin{align}
  H_{\text{eff}}^{\text{BCS}} & = \sum_{\bm{k}}[E_{\bm{k},+}\bar{\gamma}_{\bm{k},\uparrow}\gamma_{\bm{k},\uparrow} + E_{\bm{k},-}\bar{\gamma}_{-\bm{k},\downarrow}\gamma_{-\bm{k},\downarrow}]+E_g, \label{asymhopNHBCS_NHBCS_3d_diagnoalized_NHBCS_Hamiltonian_eq}
\end{align}
with
\begin{align}
  E_{\bm{k},\pm} & = \tilde{\epsilon}_{\bm{k}}\pm i\text{Im}\xi_{\bm{k}}, \label{asymhopNHBCS_NHBCS_3d_eigenvalues_definition_eq} \\
  E_g            & = E_0-\sum_{\bm{k}}E_{\bm{k},-},
\end{align}
where $\tilde{\epsilon}_{\bm{k}}=\sqrt{(\text{Re}\xi_{\bm{k}})^{2}+\Delta\bar{\Delta}}$.
The quasiparticle operators are calculated by using the Bogoliubov transformations as
\begin{align}
   & \bar{\gamma}_{\bm{k},\uparrow} = u_{\bm{k}}c_{\bm{k},\uparrow}^{\dagger}-\bar{v}_{\bm{k}}c_{-\bm{k},\downarrow},                 \\
   & \gamma_{-\bm{k},\downarrow} = v_{\bm{k}}c_{\bm{k},\uparrow}^{\dagger} + u_{\bm{k}}c_{-\bm{k},\downarrow},                        \\
   & \gamma_{\bm{k},\uparrow} = u_{\bm{k}}c_{\bm{k},\uparrow} - v_{\bm{k}}c_{-\bm{k},\downarrow}^{\dagger},                \\
   & \bar{\gamma}_{-\bm{k},\downarrow} = \bar{v}_{\bm{k}}c_{\bm{k},\uparrow} + u_{\bm{k}}c_{-\bm{k},\downarrow}^{\dagger},
\end{align}
with the coefficients
\begin{align}
   & u_{\bm{k}} = \sqrt{\frac{\tilde{\epsilon}_{\bm{k}}+\text{Re}\xi_{\bm{k}}}{2\tilde{\epsilon}_{\bm{k}}}},                                                \\
   & v_{\bm{k}} = -\sqrt{\frac{\tilde{\epsilon}_{\bm{k}}-\text{Re}\xi_{\bm{k}}}{2\tilde{\epsilon}_{\bm{k}}}}\frac{\sqrt{\Delta}}{\sqrt{\bar{\Delta}}},       \\
   & \bar{v}_{\bm{k}} = -\sqrt{\frac{\tilde{\epsilon}_{\bm{k}}-\text{Re}\xi_{\bm{k}}}{2\tilde{\epsilon}_{\bm{k}}}}\frac{\sqrt{\bar{\Delta}}}{\sqrt{\Delta}},
\end{align}
where $u_{\bm{k}}^{2}+v_{\bm{k}}\bar{v}_{\bm{k}}=1$. Although $\bar{\gamma}_{\bm{k},\sigma}$ is not the Hermitian conjugate of $\gamma_{\bm{k},\sigma}$ due to the complexity of $\Delta$ and $\bar{\Delta}$,
the quasiparticle operators satisfy the anticommutation relations
$\{\bar{\gamma}_{\bm{k},\sigma}, \gamma_{\bm{k}^{'},\sigma^{'}} \}=\delta_{\bm{k},\bm{k}'}\delta_{\sigma,\sigma'}$.
The BCS states~\eqref{asymhopNHBCS_BCS_state_Right_rep_eq} and \eqref{asymhopNHBCS_BCS_state_Left_rep_eq} satisfy
$\gamma_{\bm{k},\sigma}|\text{BCS}\rangle_{R}=0$ and $\bar{\gamma}_{\bm{k},\sigma}^{\dagger}|\text{BCS}\rangle_{L}=0$.
In the end, 
we obtain the NH gap equation as
\begin{equation}
  \frac{1}{U} = \frac{1}{N}\sum_{\bm{k}}\frac{1}{2\tilde{\epsilon}_{\bm{k}}}
  = \int d\epsilon\frac{D(\epsilon)}{2\sqrt{\epsilon^2+\Delta_0^2}},\label{asymhopNHBCS_NH_gap_eq}
\end{equation}
where the effective density of states (DOS)
is defined as
\begin{align}
  D(\epsilon) = \frac{1}{N}\sum_{\bm{k}}\delta(\epsilon-\text{Re}\xi_{\bm{k}}),\label{DOS}
\end{align}
where $\delta(x)$ is the delta function.
We note that the DOS is given only by the real part of the dispersion relations
and the imaginary part does not appear in Eq.~\eqref{DOS}.
This means that the DOS is not affected by the asymmetry of the hopping.

The gap equation~\eqref{asymhopNHBCS_NH_gap_eq} is different from that in the Hermitian system
since $U$ and $\Delta\bar{\Delta}$ are complex values.
Hereafter, we take the special gauge $\Delta_{0}=\Delta=\bar{\Delta}$ without loss of generality.
In this case, the order parameter $\Delta_0$ is complex and smoothly approaches the real value with $\gamma\rightarrow 0$.
To study the stability of the superfluid state against dissipation,
we calculate the condensation energy as
\begin{align}
  \frac{E_{\text{cond}}}{N} & = \frac{E_g-E_N}{N} \nonumber                                                                              \\
                            & =\frac{\Delta_{0}^{2}}{U} -\int d\epsilon D(\epsilon)\left(\sqrt{\epsilon^2+\Delta_0^2}-|\epsilon|\right),
  \label{asymhopNHBCS_condenasation_energy_rep_eq}
\end{align}
where $E_N$ is the energy for the normal state.
We use the real part of the condensation energy~\eqref{asymhopNHBCS_condenasation_energy_rep_eq}
to discuss the stability of the superfluid state.
The average of the filling is given by
\begin{align}
  n & =\frac{1}{N}\sum_{\bm{k},\sigma}{}_{L}\langle c_{\bm{k},\sigma}^{\dagger}c_{\bm{k},\sigma}\rangle_{R} \\
    & =\int d\epsilon D(\epsilon)\left(1-\frac{\epsilon}{\sqrt{\epsilon^2+\Delta_0^2}}\right).
\end{align}
When $\mu=0$, we get $n=1$.
As the number-preserving state is realized in realistic systems, the expectation value of the total particle number $\sum_{\bm{k},\sigma}{}_{L}\langle c_{\bm{k},\sigma}^{\dagger} c_{\bm{k},\sigma}\rangle_{R}$ is conserved and gives a real number. Hereafter, we set $\mu=0$.

\begin{figure}[bt]
  \centering
  \includegraphics[width=7.5cm]{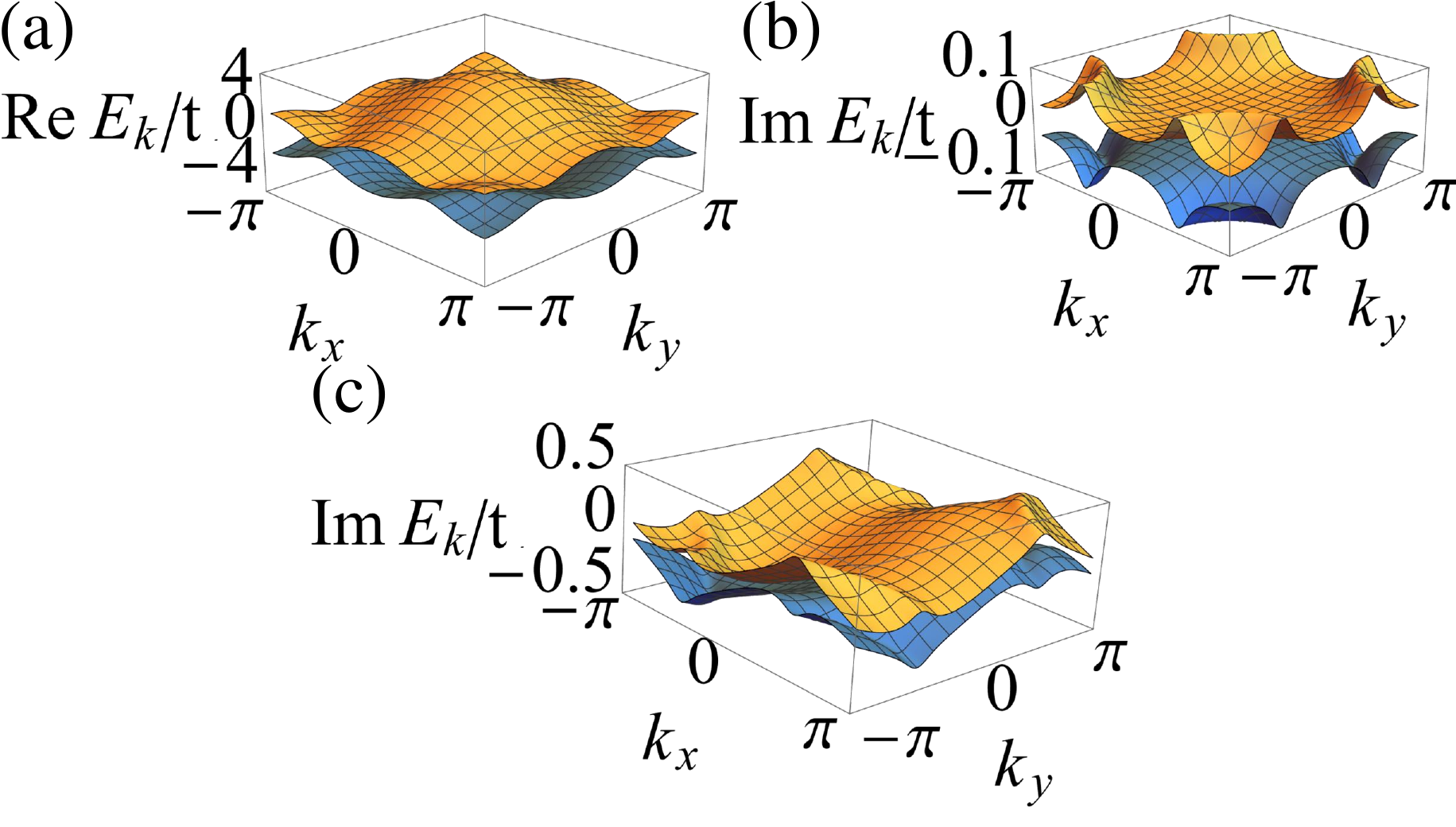}
  \caption{
    (a) The real part and (b) the imaginary part of the quasiparticle energy $E_{\bm{k}}=E_{\bm{k},+},-E_{\bm{k},-}$ (orange for the positive sign and blue for the negative sign) when $\delta=0$.
    (c) The imaginary part of the $E_{\bm{k}}$ when $\delta=0.1$. Dispersion relations with $k_z=0$
    for quasiparticles in the system with $U_1/t=3$ and $\gamma/t=1$ are used.
  } %
  \label{asymhopNHBCS_ReImEK_image}
\end{figure}

We note that the effect of the asymmetric hopping $\delta$ does not appear in the effective DOS~\eqref{DOS},
the NH gap equation~\eqref{asymhopNHBCS_NH_gap_eq},
and the condensation energy~\eqref{asymhopNHBCS_condenasation_energy_rep_eq}.
This means that the asymmetric hopping has no effect on the superfluidity within the mean-field approximation.
In particular, when $\gamma=0$, the conventional superfluid state with a finite order parameter is stable against
the asymmetric hopping within the mean-field approximation
although its lifetime should be finite.
By contrast, the quasiparticle energy $E_k$ depends on $\delta$.
Here, we show in Fig.~\ref{asymhopNHBCS_ReImEK_image} the real and imaginary parts of the quasiparticle energy $E_{\bm{k}}$
in the system with $U_1/t=3$ and $\gamma/t=1$
when $\delta=0$ and $0.1$.
The real part of the quasiparticle energy is regarded as the effective energy of quasiparticles and the imaginary part of that is regarded as the lifetime of quasiparticles.
When we increase $\delta$, the real part of the quasiparticle energy never changes,
while the imaginary part is induced with changing the sign in reciprocal space, as shown in Fig.~\ref{asymhopNHBCS_ReImEK_image}(c).
This means that the lifetime of the quasiparticles on particular regions is amplified.
If the asymmetric hopping is considered in the Hamiltonian as
$H=-t\sum_{\langle i,j \rangle,\sigma} (e^{\alpha} c_{i,\sigma}^{\dagger}c_{j,\sigma} + e^{-\alpha}c_{j,\sigma}^{\dagger}c_{i,\sigma})$, both $t$ and $\delta$
in our Hamiltonian \eqref{asymhopNHBCS_effHamltonian_real_space_rep_eq} are formally changed.
However, by rescaling the energy unit, we obtain
the simple phase diagram, which will be discussed
in Appendix~\ref{app_other_asymmetric_hopping}.

\section{Generalization to arbitrary dimensions} \label{sec_generalization_for_d_dim}
\begin{figure}[htb]
  \centering
  \includegraphics[width = 7.5cm]{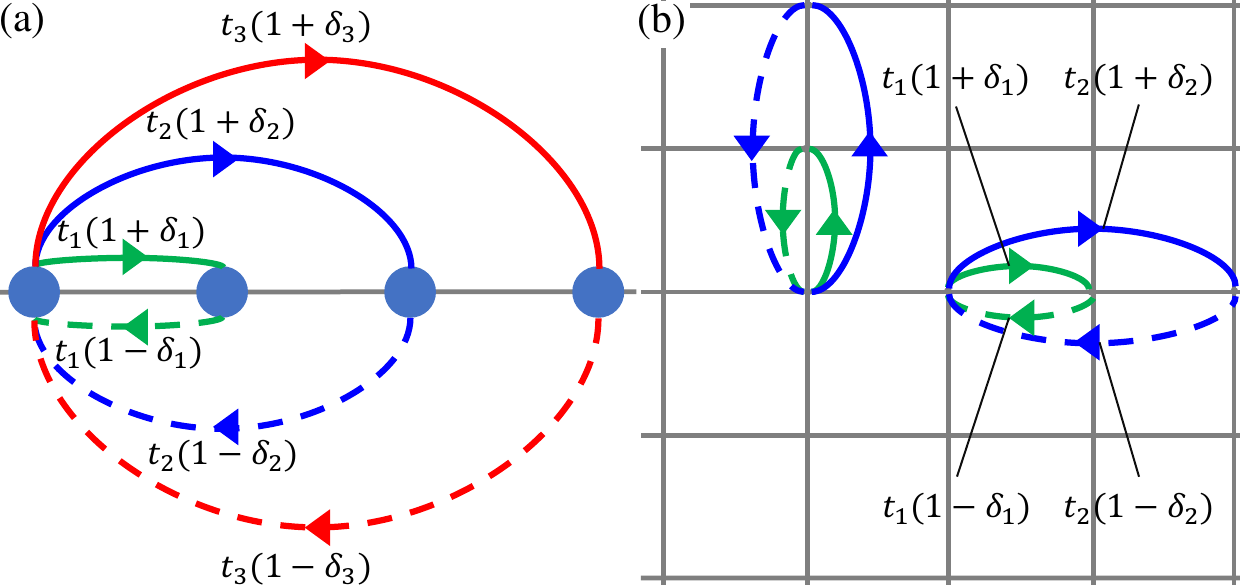}
  \caption{\cred{Schematic image of the asymmetric hopping in the $d$-dimensional hypercubic lattice when (a) $d=1$ and (b) $d=2$. We consider the hopping across several lattice sites in the same direction, for example, the hopping amplitude between nearest neighbor sites is $t_{1}(1\pm \delta_{1})$, the hopping between next-nearest neighbor sites is $t_{2}(1\pm \delta_{2})$, and the hopping between $m$-th nearest neighbor sites is $t_{m}(1\pm \delta_{m})$.}}
  \label{asymhopNHBCS_d_dim_asymhop_1d2d_image}
\end{figure}
In this section, we consider a $d$-dimensional NH Hubbard model
with asymmetric hopping and complex-valued interactions.
%
For the sake of simplicity, we consider a hypercubic lattice.
The Hamiltonian is given by
\begin{align}
  H_{\text{eff}} & = H_{\text{kin}} + H_{\text{int}},\label{gen}                                                                                       \\
  H_{\text{kin}} & = -\sum_m t_{m}\sum_{\langle i,j\rangle_m ,\sigma}\left[(1+\delta_{m})c_{i,\sigma}^{\dagger}c_{j,\sigma}\right. \notag \\
  & \left.+(1-\delta_{m})c_{j,\sigma}^{\dagger}c_{i,\sigma}\right], \label{asymhopNHBCS_kinetic_term_on_d_dimensional_eq}       \\
  H_{\text{int}} & = -U\sum_{i}c_{i,\uparrow}^{\dagger}c_{i,\downarrow}^{\dagger}c_{i,\downarrow}c_{i,\uparrow},
\end{align}
\cred{where $t_{m}$ is the hopping to the $m$-th nearest neighbor sites in the same direction, $\langle i,j\rangle_m$ means the summation over the $m$-th nearest neighbor pairs, and $\delta_m$ is its asymmetry. In our discussions, we only consider the hopping across lattice sites in the same direction (see Fig.~\ref{asymhopNHBCS_d_dim_asymhop_1d2d_image}),
which is regarded as the generalization of the Hatano-Nelson model~\cite{hatano96,hatano97,hatano98}.}
In the noninteracting case, the dispersion relation is given as
\begin{align}
  \xi_{\bm{k}} & = -\sum_{m}\sum_{l=1}^{d}\Big[ 2t_{m}\cos (mk_{l})
  +2it_{m}\delta_{m}\sin (mk_{l}) \Big].
  \label{asymhopNHBCS_Bloch_Hamiltonian_on_d_dimensional_tight_bindind_eq}
\end{align}
The long-range hopping and dimensionality in the system can be captured
in Eq.~\eqref{asymhopNHBCS_Bloch_Hamiltonian_on_d_dimensional_tight_bindind_eq},
and the asymmetric effect only appears in its imaginary part.
This property is essentially the same as the simpler case discussed above.
Therefore, no effect of the asymmetric hopping appears, as far as we consider the generalized model~\eqref{gen}
in the framework of the BCS theory.

\begin{figure}[bt]
  \centering
  \includegraphics[width=\linewidth]{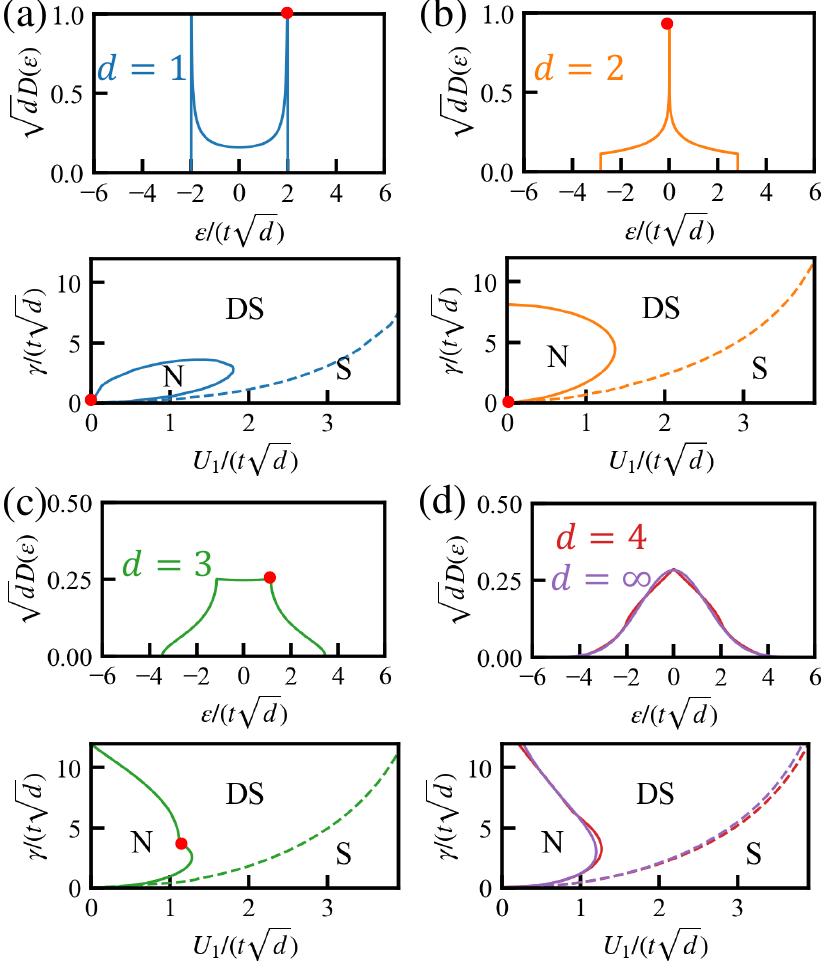}
  \caption{The DOS and the phase diagram of the $d$-dimensional attractive Hubbard model with two-body loss and asymmetric nearest neighbor hopping when (a) $d=1$, (b) $d=2$, (c) $d=3$, and (d) $d=4,\infty$. N, DS, and S denote the normal state, the DS state, and the superfluid state, respectively. The hopping amplitude is rescaled as $t\sqrt{d}$ based on the energy unit on an infinite dimensional system~\cite{muller89,metzner89}. The red point denotes the singular point in DOS in upper figures, and it affects the phase boundary in lower figures.}
  \label{asymhopNHBCS_PD_for_d_dimensions_image}
\end{figure}

Figure~\ref{asymhopNHBCS_PD_for_d_dimensions_image} shows the DOS and the phase diagram
on the $d$-dimensional NH Hubbard model with nearest-neighbor hopping.
\cred{
The normal (N) state is characterized by a trivial solution of the NH gap equation~\eqref{asymhopNHBCS_NH_gap_eq}, and the superfluid (S) and the dissipation-induced superfluid (DS) states are characterized by the nontrivial solution. In the S (DS) state, the real part of the condensation energy is $\text{Re}E_{\text{cond}}<0 (>0)$.
}
We find that, in the phase diagram, the N state appears in the small $U$ and intermediate dissipation rate $\gamma$,
the ordered state with finite $\Delta_0$ is widely realized for any dimensions.

One of the important features inherent in the NH system is the emergence of the exceptional points (EPs).
When the system crosses the phase boundary from the DS state to the N state,
the real part of the order parameter gradually approaches to zero, while the imaginary part suddenly vanishes from $\epsilon_0$ which satisfies $\epsilon_{0} = \pm \text{Im}\Delta_{0}/t$.
When the transition occurs, the effective Hamiltonian cannot be diagonalized, and EPs emerge in the reciprocal space.
This is equivalent to the divergence of the integral function at $\epsilon=\epsilon_0$ in the gap equation~\eqref{asymhopNHBCS_NH_gap_eq}.
Importantly, this means that EPs as well as DOS play a crucial role in the NH gap equation~\cite{takemori24}.
\cred{Namely, the DS state is expanded due to the large DOS at $\epsilon=\epsilon_{0}$.}
In one dimensions, the N state is surrounded by the DS state
in the phase diagram,
which is shown in Fig.~\ref{asymhopNHBCS_PD_for_d_dimensions_image}(a).
The corresponding phase boundary strongly reflects
the singularity in the DOS at $\epsilon/t=\pm 2$.
This behavior is characteristic of one dimensions.
In two dimensions, DOS has a singularity at $\epsilon=0$.
However, this little affects the phase diagram
since the corresponding phase boundary is located
at the origin of the diagram,
as shown in Fig.~\ref{asymhopNHBCS_PD_for_d_dimensions_image}(b).
%
%
The DOS in three dimensions has the cusp singularity at $\epsilon/t=2$,
leading to the small cusp singularity in the phase boundary at $(U_{1}/t,\gamma/t)\sim(1.9,6.5)$.
In larger dimensions, the singularity in the DOS tends to smear, and
no singularity appears in the hypercubic lattice with $d\rightarrow\infty$.

These results indicate that the reentrant superfluidity for small $U_1$ with increasing dissipation is ubiquitous for any dimensional hypercubic lattices.
%
Such a unique phase boundary can be detected by using the photoassociation techniques~\cite{tomita19} and postselecting the null measurement outcomes with the use of the quantum-gas microscope~\cite{ashida16,ott16,mitra18,brown20,chan20,hartke23}.


\begin{figure}[h]
  \centering
  \includegraphics[width=\linewidth]{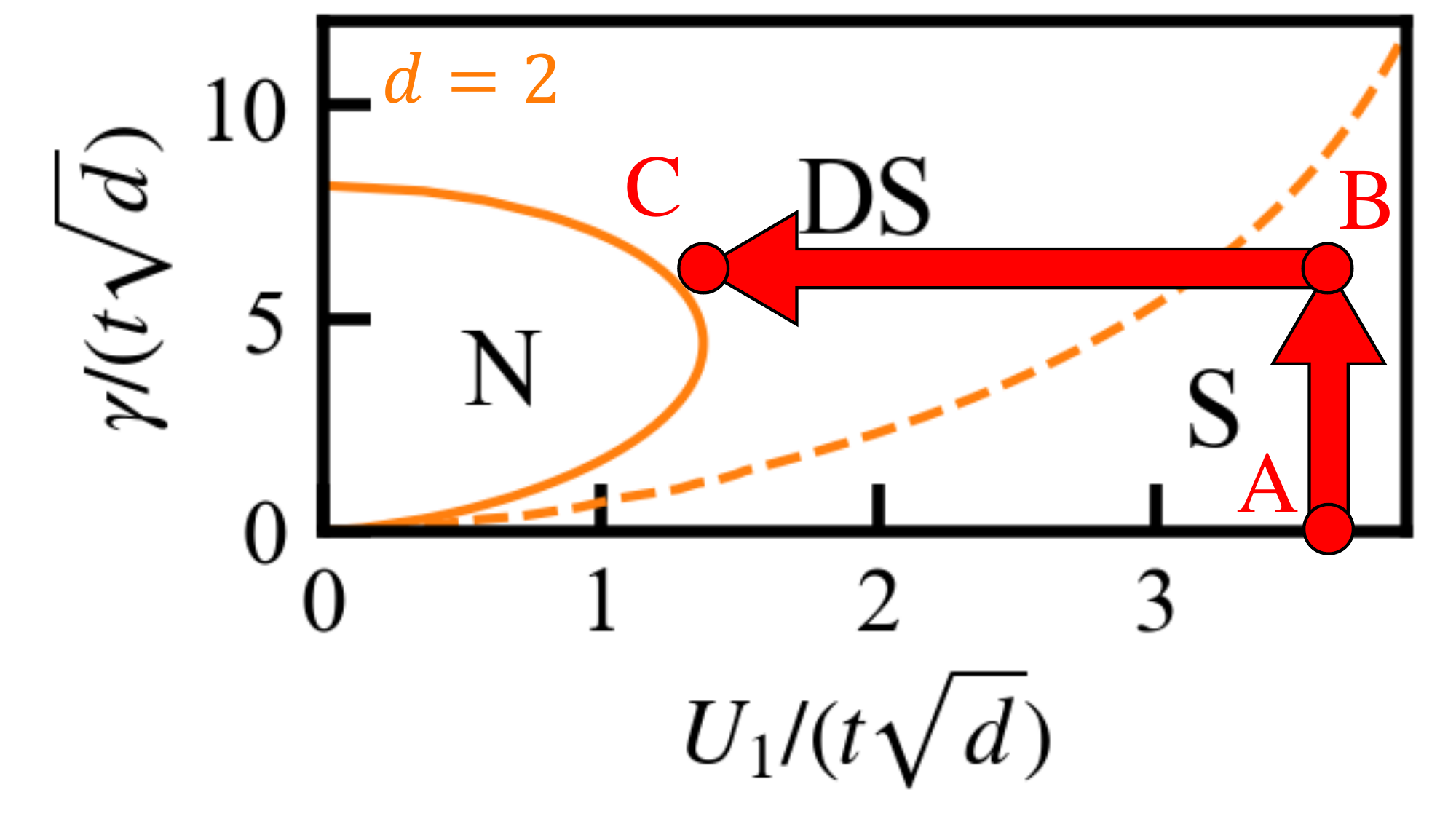}
  \caption{\cred{Experimental protocol to reach the dissipation-induced superfluid for strong dissipation. Here, we use the phase diagram for the square lattice and the same protocol can be done in the other dimensions. First, we prepare a superfluid state for large attraction $U_{1}$ (A) and then introduce large dissipation $\gamma$ by utilizing a photoassociation technique (B). Finally, the system will reach the dissipation-induced superfluid regime by decreasing $U_{1}$ (C).}}
  \label{asymhopNHBCS_protocols_DS_and_normal_image}
\end{figure}

\section{Conclusions and discussions}
\label{sec_discussion}
In this paper, we have considered the three-dimensional attractive Hubbard model
with fundamental dissipation, i.e., asymmetric hopping and complex-valued interactions
which can be realized by the collective one-body loss and two-body loss in the ultracold fermionic systems.
We have demonstrated that the asymmetry of the hopping has no effects on
the fermionic superfluidity within the NH BCS approximation.
It has been found that the complex-valued interactions yield the reentrant phase transitions,
as discussed in the previous works.
In particular, we have confirmed that the singularity of the noninteracting DOS
leads to the singularity in the phase boundary
between the dissipation-induced superfluid and normal states.

Our system can be implemented in ultracold atoms~\cite{yamamoto19}.
\cred{
  We can use, e.g., $^{6}\mathrm{Li}$, $^{40}\mathrm{K}$ and $^{173}\mathrm{Yb}$ in an optical lattice with attractive interaction, two-body loss, and asymmetric hopping. The hopping amplitude $t$ can be changed by tuning the lattice depth. By using the Feshbach resonance, we can control the attractive interaction $U_{1}$~\cite{chin10}. The asymmetric hopping can be realized by means of the nonlocal Rabi coupling~\cite{gong18,liu19,he21} and two-body loss can be realized by using the photoassociation~\cite{tomita17,honda23}. For the detailed discussion, we refer the reader to Appendix~\ref{app_implementation_asymhop}.
  }
\cred{The NH phase for strong dissipation can be accessed in experiments by using the QZE. As the timescale of quantum jumps is estimated as $1/\gamma$ and the relaxation timescale towards the quasiequilibrium state is given by $1/t$, the energy scale of $\gamma$ and $t$ should be comparable to realize the NH dynamics. In Fig.~\ref{asymhopNHBCS_protocols_DS_and_normal_image}, we illustrate the experimental protocol for reaching the NH phase in the strong dissipation regime that satisfies $\gamma/t\sim 5$. We first prepare a superfluid for large attraction $U_{1}$ [(A) in Fig.~\ref{asymhopNHBCS_protocols_DS_and_normal_image}] and increase the two-body loss rate [(B) in Fig.~\ref{asymhopNHBCS_protocols_DS_and_normal_image}] by using photoassociation techniques~\cite{tomita17,honda23}. Finally, we ramp down the attraction $U_{1}$ with the use of the Feshbach resonance [(C) in Fig.~\ref{asymhopNHBCS_protocols_DS_and_normal_image}]. If we do not have dissipation, atoms that form the on-site molecular pairs tunnel to neighboring sites at a timescale $1/t$. However, under strong dissipation, the hopping of the atoms to the neighbor sites are suppressed due to QZE and the atoms are tightly coupled to form on-site molecules. Then, we can see the delay of dissociation of such QZE-assisted molecules even after a timescale $1/t$, and this can be a signature of the dissipation-induced superfluid.}

Although we consider the hypercubic lattice, it is interesting to explore the NH quantum phase transition of the dissipative Hubbard model with asymmetric hopping on a bipartite lattice. The asymmetric hopping in NH systems will play a crucial role in studying the symmetry and topological classification~\cite{gong18,kawabata19,kawabata22}. Then, exploring the topological phase transition in the NH Hubbard model with asymmetric hopping may be a novel research. Furthermore, investigating the attractive NH Hubbard model on the one-dimensional system with multiple dissipation with the use of a more precise method, such as the Bethe ansatz~\cite{fukui98,nakagawa21}, remains future work. Analysis of the Lindblad dynamics with asymmetric hopping and two-body loss is another important pursuit. We hope that this paper contributes to the understanding of the NH system with fundamental dissipation. In the view of quantum trajectory approach~\cite{daley14}, the NH system describes the special case of the dynamics of the quantum trajectory which represents the measurement-induced phase transition~\cite{fuji20,goto20,tang20,jian21SYK,block22,minato22,doggen22,yamamoto23local,zhou24}. More recently, as the dissipative system with gain has been realized in ultracold atoms~\cite{tsuno24}, the study of the NH fermionic superfluidity with gain and loss is also interesting. Exploring the NH many-body physics can give a fundamental concept of open quantum systems.

\begin{acknowledgments}
  We would like to thank Hiroyuki Tajima and Masaya Nakagawa for fruitful discussions.
  This work was supported by Grant-in-Aid for Scientific Research from JSPS,
  KAKENHI Grants No. JP23K19031 (K.Y.), and No. JP22K03525 (A.K.).
  K.Y. was also supported by Yamaguchi Educational and Scholarship Foundation, Toyota RIKEN Scholar program, Murata Science and Education Foundation, and Public Promoting Association Kura Foundation.
\end{acknowledgments}

\appendix

\begin{figure}[htb]
  \centering
  \includegraphics[width =\linewidth]{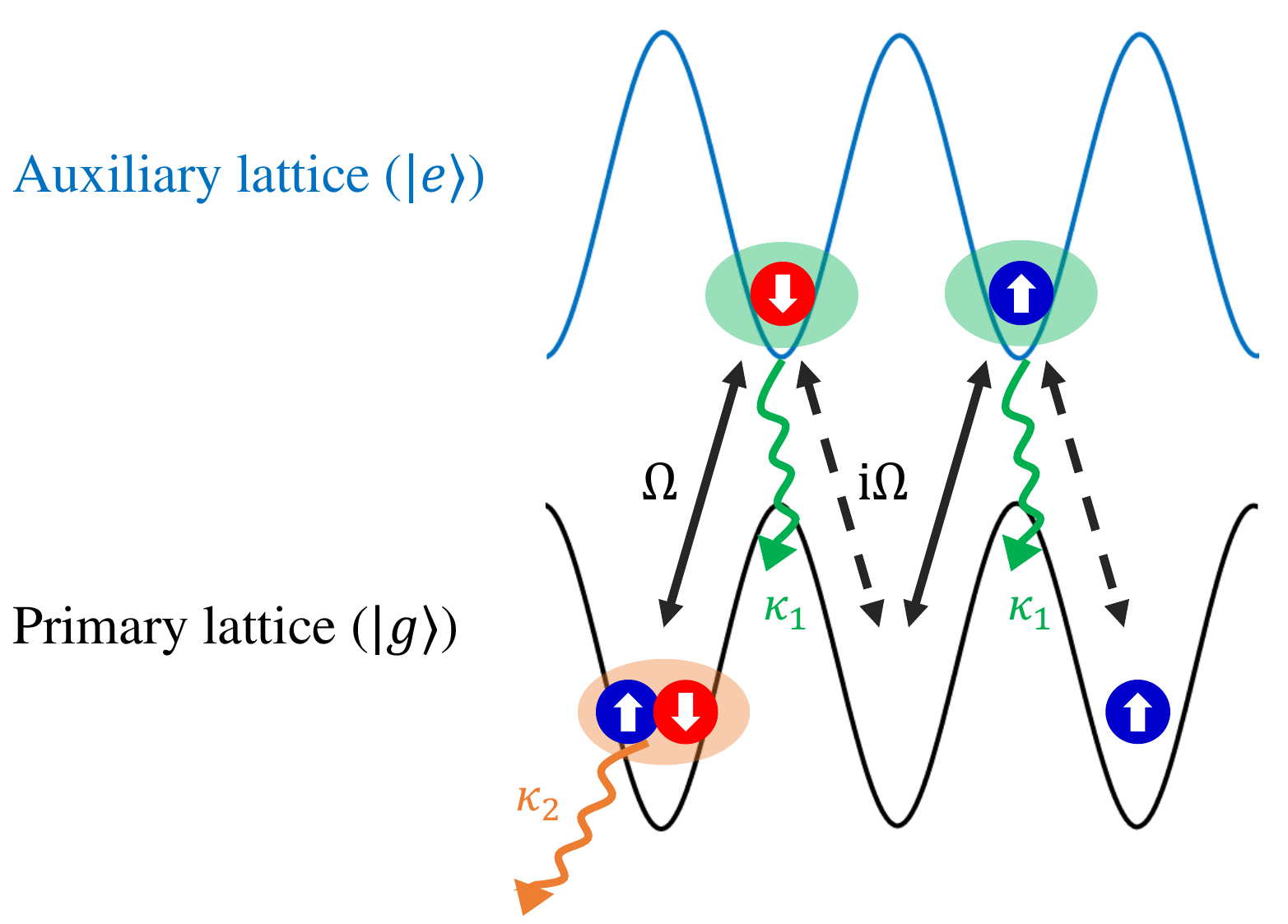}
  \caption{\cred{Schematic image of our system. We consider the dissipative dynamics of fermionic atoms in primary (Black) and auxiliary (Blue) lattices. The Rabi coupling between the primary and auxiliary lattices is introduced by using a running wave~\cite{gong18} and the rate is changed by $i$ compared with the left nearest sites. We introduce the one-body loss with rate $\kappa_{1}$ to the auxiliary lattice and the two-body loss with rate $\kappa_{2}$ to the primary lattice.}}
  \label{asymhopNHBCS_primary_auxiliary_optlat_image}
\end{figure}
\begin{figure*}[tb]
  \centering
  \includegraphics[width=\textwidth]{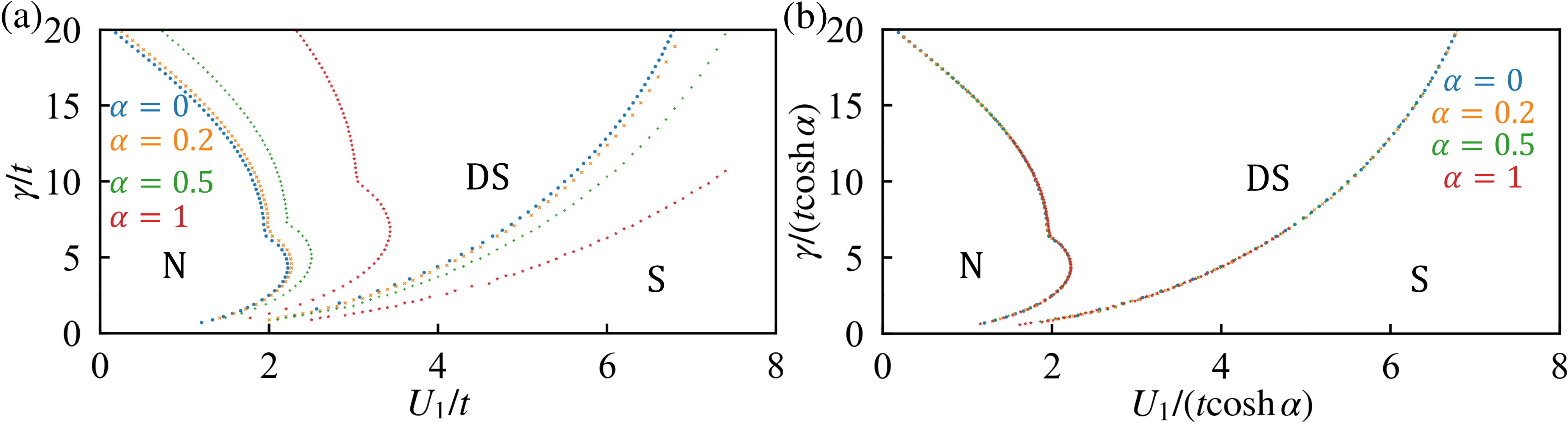}
  \caption{Phase diagram of the NH BCS model with asymmetric hopping and complex-valued interactions with energy unit (a) $t$ and (b) $t\cosh\alpha$. N denotes the normal state where the gap equation~\eqref{asymhopNHBCS_NH_gap_eq} has a trivial solution. The gap equation has a non-trivial solution and $\text{Re}E_{\text{cond}}<0$ in the superfluid state, which is denoted as S. DS state appears when a non-trivial solution of the NH gap equation exists and $\text{Re}E_{\text{cond}}>0$. In Fig.~\ref{asymhopNHBCS_PD_alphas_image}(a), as increasing $\alpha$, the phase boundaries shift to the lower right. In Fig.~\ref{asymhopNHBCS_PD_alphas_image}(b), By rescaling the energy unit, the phase boundary for different $\alpha$ collapses into the single phase boundary.}
  \label{asymhopNHBCS_PD_alphas_image}
\end{figure*}
\section{\cred{Experimental setup for realizing our system}} \label{app_implementation_asymhop}
\cred{
  In this section, we briefly explain how the asymmetric hopping can be realized in experiments,
  following the paper~\cite{gong18}. We start from the ultracold fermionic atoms in an optical lattice with an auxiliary lattice, where the potential minima of the auxiliary lattice are positioned at the middles for that of the primary lattice, as shown in Fig.~\ref{asymhopNHBCS_primary_auxiliary_optlat_image}. For the sake of simplicity, we treat with a one-dimensional system, and extension for higher dimensions will follow similarly.
} The full dynamics is described by the following Markovian Lindblad equation~\cite{daley14}
\begin{equation}
  \frac{\partial \rho}{\partial t} = -i[H_{0}+V,\rho] +  \mathcal{D}[\tilde{L}_{i,\sigma}^{(1)}]\rho + \mathcal{D}[\tilde{L}_{i}^{(2)}]\rho,
\end{equation}
where $\rho$ is the density matrix. Here, the Hermitian Hamiltonian $H=H_{g}+H_{e}+V$ is written as
\begin{align}
  & H_{g} = -t \sum_{\langle i,j\rangle,\sigma} (c_{i,\sigma,g}^{\dagger}c_{j,\sigma,g} +\text{H.c.}) - \mu\sum_{i,\sigma}n_{i,\sigma,g} \notag \\
  &-U_{1}\sum_{i}c_{i,\uparrow,g}^{\dagger}c_{i,\downarrow,g}^{\dagger}c_{i,\downarrow,g}c_{i,\uparrow,g}, \label{asymhopNHBCS_Hermitian_attractive_Hubbard_Hamiltonian_eq} \\
  & H_{e} = -t \sum_{\langle i,j\rangle,\sigma} (c_{i,\sigma,e}^{\dagger}c_{j,\sigma,e} +\text{H.c.}) - \mu\sum_{i,\sigma}n_{i,\sigma,e}, \\
  & V  = \frac{\Omega}{2}\sum_{i,\sigma} [c_{i,\sigma,e}^{\dagger}(c_{i,\sigma,g}+ic_{i+1,\sigma,g}) + \text{H.c.}],
\end{align}
where $c_{i,\sigma,\alpha}(c_{i,\sigma,\alpha}^{\dagger})$ is the annihilation (creation) operator of the fermion with spin $\sigma=\uparrow,\downarrow$ at site $i$. The primary (auxiliary) lattice is denoted as $\alpha=g (e)$ and the coupling strength between the primary and auxiliary lattices is introduced as $\Omega$. To describe the dissipation, we introduce the Lindblad operator $\tilde{L}_{m}$ that describes the dissipative event at characteristic rates. \cred{The one-body loss of the auxiliary lattice is described by $\tilde{L}_{i,\sigma}^{(1)}=\sqrt{\kappa_{1}}c_{i,\sigma,e}$ where $\kappa_{1}$ is the dissipation rate}. The operator $\tilde{L}_{i}^{(2)}=\sqrt{\kappa_{2}}c_{i,\downarrow,g}c_{i,\uparrow,g}$ describes the two-body loss of the primary lattice at site $i$ at a rate $\kappa_{2}$, which generates the complex-valued interaction.

To extract the effective dynamics of the primary lattice, we employ the adiabatically elimination~\cite{reiter12,gong18,liu19,he21}, which changes the local loss of the auxiliary lattice into the nonlocal loss of the primary lattice. Then, \cred{in the regime where $\kappa_{1}\gg \Omega$}, the Lindblad equation which describes the effective dynamics of the primary lattice is written as
\begin{equation}
  \frac{\partial\rho}{\partial t} = -i[H_{g},\rho] + \mathcal{D}[L_{\text{eff},i,\sigma}^{(1)} ]\rho + \mathcal{D}[L_{\text{eff},i}^{(2)}]\rho. \label{asymhopNHBCS_effective_lindblad_adiabatically_eliminated_eq}
\end{equation}
The effective Lindblad operator $L_{\text{eff},i,\sigma}^{(1)},L_{\text{eff},i}^{(2)}$ are the same as in Eq.~\eqref{asymhopNHBCS_effective_lindblad_eq}, where the parameters are related to each other by
\begin{equation}
  \gamma_{1}=\Omega^{2}/\kappa_{1},\ \gamma_{2}=\kappa_{2}.
\end{equation}
Then, by focusing on the effective short-time dynamics that is described by the NH Hamiltonian obtained from Eq.~\eqref{asymhopNHBCS_effective_lindblad_adiabatically_eliminated_eq}, we arrive at Eq.~\eqref{asymhopNHBCS_effHamltonian_real_space_rep_eq} in Sec.~\ref{sec_model}.

\cred{
  Here, we present the detailed values of experimental parameters by using $^{173}\mathrm{Yb}$ as an example. First, the hopping amplitude $t$ can be changed by tuning the lattice depth. By using the Feshbach resonance, we can control the attractive interaction $U_{1}$. In Ref.~\cite{honda23}, the hopping amplitude and repulsive interaction strength are estimated to be several $\mathrm{kHz}$ for the repulsive Fermi Hubbard model on the dimerized lattice. By tuning the magnetic field strength, we can change the interaction strength to be negative. Second, we can control the one-body loss rate $\kappa_{1}$ in the auxiliary lattice and the asymmetry of the hopping $\delta$ by tuning the Rabi coupling $\Omega$ between the primary and auxiliary lattices. The coupling constant between two hyperfine state is estimated to be several $\mathrm{kHz}$~\cite{ren22}. The one-body loss can be introduced by using a near-resonant optical beam, e.g., the loss rate $\kappa_{1}\sim \mathrm{kHz}$ is used with $^{6}\mathrm{Li}$ in Ref.~\cite{corman19}. Similar protocols will follow in $^{173}\mathrm{Yb}$ and it is necessary to enhance the loss rate by employing a high power beam to realize our model. For example, if we use $\Omega\sim 2\pi \times 6.3 \mathrm{kHz}$ and $t\sim 2\pi\times 0.32 \mathrm{kHz}$ given in Refs.~\cite{ren22,honda23}, the one-body loss rate $\kappa_{1} = \Omega^{2}/(2t\delta)\sim 2\pi \times 3.1\times 10^{2} \mathrm{kHz}$ should be required, and the asymmetry of the hopping is given by $\delta = 0.2$ in this case. Third, the implementation of the two-body loss can be done by using the photoassociation~\cite{tomita17}. For $^{173}\mathrm{Yb}$, the two-body loss rate is tunable up to several $\mathrm{ms}^{-1}$~\cite{honda23}, which means that $\gamma_{2}/t\sim 5$ can be reached. Finally, to extract the conditional dynamics where the quantum jump does not occur, we can employ the quantum-trajectory method and obtain the special measurement outcome by using the quantum-gas microscope~\cite{ashida16,ott16,mitra18,brown20,chan20,hartke23}. Although the quantum-gas microscope is available in one- and two-dimensional systems, the NH phase in three-dimensional systems can be observed by focusing on the short time dynamics. Thus, the parameter region of our model can be accessed in the experiment.
}

\section{The effect of the other type of the asymmetric hopping} \label{app_other_asymmetric_hopping}
Here we consider the other type of the asymmetric hopping on a three-dimensional Hubbard model with a complex-valued interaction. The kinetic term of the NH Hamiltonian is given by
\begin{equation}
  H_{\text{kin}} = -t'\sum_{\langle i,j \rangle,\sigma} (e^{\alpha} c_{i,\sigma}^{\dagger}c_{j,\sigma} + e^{-\alpha}c_{j,\sigma}^{\dagger}c_{i,\sigma}), \label{asymhopNHBCS_other_type_asymmetric_hopping_kinetic_term_eq}
\end{equation}
where the asymmetry of the hopping is $\alpha$. After the Fourier transformation, the energy dispersion is written as
\begin{equation}
  \xi_{\bm{k}} = -2t'\cosh\alpha\sum_{j=x,y,z}\cos k_{j} -2it'\sinh\alpha \sum_{j=x,y,z}\sin k_{j}. \label{asymhopNHBCS_other_type_asymmetric_hopping_energy_dispersion_eq}
\end{equation}
Then, the NH BCS Hamiltonian~\eqref{asymhopNHBCS_NHBCS_effective_Hamiltonian_rep_eq}, the NH gap equation~\eqref{asymhopNHBCS_NH_gap_eq}, and the condensation energy~\eqref{asymhopNHBCS_condenasation_energy_rep_eq} are given by replacing $\xi_{\bm{k}}$ by that in Eq.~\eqref{asymhopNHBCS_other_type_asymmetric_hopping_energy_dispersion_eq}. In Fig.~\ref{asymhopNHBCS_PD_alphas_image}, we show the phase diagram of the NH BCS model with complex-valued interactions and the asymmetric hopping in Eq.~\eqref{asymhopNHBCS_other_type_asymmetric_hopping_kinetic_term_eq}. Here, we discuss how the asymmetric hopping affects the phase diagram. As increasing the asymmetric hopping amplitude $\alpha$, the phase boundary between the normal and the DS state shifts to the upper right [see the lines for $\alpha=0.2,0.5,\text{and} 1.0$ in Fig.~\ref{asymhopNHBCS_PD_alphas_image}(a)]. The phase boundary between the DS state and the superfluid state also shifts to the lower right in Fig.~\ref{asymhopNHBCS_PD_alphas_image}(a). These shifts of the phase boundary indicate that the superfluidity becomes unstable due to the asymmetry of the hopping $\alpha$. However, the phase boundary with different asymmetric hopping collapses into a single phase boundary [see Fig.~\ref{asymhopNHBCS_PD_alphas_image}(b)] by rescaling the energy unit as $t\to t\cosh\alpha$. To understand that, we first consider the NH gap equation~\eqref{asymhopNHBCS_NH_gap_eq}. By changing the energy unit, the NH gap equation~\eqref{asymhopNHBCS_NH_gap_eq} for $\alpha>0$ is the same as that for $\alpha=0$. Thus, the phase diagram for different asymmetry of the hopping $\alpha$ collapses into that for $\alpha=0$. Such a collapse into a single phase boundary indicates that the superfluid order parameter $\Delta_{0}$ for $\alpha>0$, and $(U_{1}/t,\gamma/t)=(U_{0}/(t\cosh \alpha),\gamma_{0}/(t\cosh \alpha))$ is $\cosh\alpha$ times than that for $\alpha=0$, and $(U_{1}/t,\gamma/t)=(U_{0}/t,\gamma_{0}/t)$. The collapse into a single line of the condensation energy $E_{\text{cond}}$ also occurs by performing the same rescaling. We can illustrate the phase diagram for complex-valued interactions with any asymmetric hopping $\alpha$ by changing the energy unit. As increasing $\alpha$, we rescale the energy unit as a larger one, which indicates that the real and the imaginary part of the order parameter and the condensation energy become small. This corresponds to the instability of the superfluids.

\end{document}